\documentclass[fleqn,usenatbib]{mnras}

\usepackage[T1]{fontenc}

\DeclareRobustCommand{\VAN}[3]{#2}
\let\VANthebibliography\thebibliography
\def\thebibliography{\DeclareRobustCommand{\VAN}[3]{##3}\VANthebibliography}

\usepackage{graphicx}	
\graphicspath{{Figures_pdf/}}
\usepackage{amsmath}	
\usepackage{amssymb}	
\usepackage{newtxtext,newtxmath}

\title[Heliospheric forcing by the solar dynamo]{Long-term forcing of the Sun's coronal field, open flux and cosmic ray modulation potential during grand minima, maxima and regular activity phases by the solar dynamo mechanism}

\author[Dash et al.]{
Soumyaranjan Dash$^{1}$, Dibyendu Nandy$^{1,2}$\thanks{E-mail:dnandi@iiserkol.ac.in}, Ilya Usoskin$^{3}$
\\
$^{1}$Center of Excellence in Space Sciences India, Indian Institute of Science Education and Research Kolkata, Mohanpur 741246, West Bengal, India\\
$^{2}$Department of Physical Sciences, Indian Institute of Science Education and Research Kolkata, Mohanpur 741246, West Bengal, India\\
$^{3}$Space Physics and Astronomy Research Unit and Sodankyl$\ddot{a}$ Geophysical Observatory, University of Oulu, 90014, Finland
}

\begin{document}
\label{firstpage}
\pagerange{\pageref{firstpage}--\pageref{lastpage}}
\maketitle

\begin{abstract}
Magnetic fields generated in the Sun's interior by the dynamo mechanism drive solar activity over a range of time-scales. Direct sunspot observations exist for few centuries, reconstructed variations based on cosmogenic isotopes in the solar open flux and cosmic ray flux exist over thousands of years. While such reconstructions indicate the presence of extreme solar activity fluctuations in the past, causal links between millennia scale dynamo activity, consequent coronal field, solar wind, open flux and cosmic ray flux variations remain elusive; lack of coronal field observations compound this issue. By utilizing a stochastically forced solar dynamo model and potential field source surface extrapolation we perform long-term simulations to illuminate how dynamo generated magnetic fields govern the structure of the solar corona and the state of the heliosphere -- as indicated by variations in the open flux and cosmic ray modulation potential. We establish differences in the nature of the large-scale structuring of the solar corona during grand maximum, minimum, and regular solar activity phases and simulate how the open flux and cosmic ray modulation potential vary across these different phases of activity. We demonstrate that the power spectrum of simulated and observationally reconstructed solar open flux time series are consistent with each other. Our study provides the theoretical foundation for interpreting long-term solar cycle variations inferred from cosmogenic isotope based reconstructions and establishes causality between solar internal variations to the forcing of the state of the heliosphere.
\end{abstract}

\begin{keywords}
Sun: activity -- Sun: corona -- Sun: heliosphere -- Sun: magnetic fields -- dynamo.
\end{keywords}



\section{Introduction}
The variability of solar magnetic activity over long time scales is manifested in multiple observable proxies. Direct solar observations for the past $\sim$ 400 years have revealed significant variability in the solar magnetic cycle, covering a period of extremely low activity known as the Maunder minimum \citep[second half of the 17th century,][]{Eddy1988ssgv} to a period of increased activity known as the Modern grand maximum \citep[middle of the 20th century][]{Usoskin2004}. Modulations in solar cycle amplitude and duration are indicators of such fluctuations. The impact of solar magnetic fields on the state of the heliosphere -- as it emerges through the surface, evolves and extends into the solar corona -- is manifested via the open solar flux (here after OSF). It is the distribution of the coronal magnetic fields that determine the solar open flux. Magnetic activity evolution of the Sun and other stars directly influences the environment of the harboured planets \citep{Nandy2004SoPh,Nandy2007AdSpR,Das2019ApJ}. Our current understanding of the long-term solar variability and its impact on solar system planets with observations, reconstructions and theoretical modelling has improved over the years \citep{Nandy2021PEPS}. However, the causal link between millennia-scale solar dynamo activity and the consequent variations in the state of the heliosphere via coronal magnetic field dynamics remains unexplored. 

Multi-millennial reconstructions of solar activity based on cosmogenic isotopes such as $^{10}$Be (in polar ice cores) and $^{14}$C (in tree rings) show the presence of grand minima, maxima and regular activity phases \citep{Usoskin2003,Usoskin2004,Usoskin2017,Wu2018}. 
The flux of galactic cosmic rays near Earth is modulated by variations in the heliospheric magnetic field which is a result of solar magnetic activity governed by the solar dynamo mechanism. Fluctuations in the cosmic ray flux provide critical insights towards understanding solar activity variations \citep{Owens2012GeoRL}. One of the critical questions in this regard is whether grand minima and maxima episodes are the outcome of special states of the solar dynamo mechanism or if they result from random variability \citep{Carbonell1994,Nandy2011,Choudhuri2012,Tripathi2021}. Direct observations of the emergence of sunspots on the solar surface which only exists from the early 17th century onwards, are not sufficient to definitively constrain such extreme episodes. Long-term solar dynamo simulations which generate such variations hold the key to explain the evolution of the solar coronal magnetic field configuration that modulates the heliospheric parameters e.g. open flux, solar wind and cosmic ray modulation potential.   

Long-term solar dynamo simulations with stochastic fluctuations in the poloidal-field source can generate extreme fluctuations with grand maximum and minimum-like episodes \citep{Pasos2014,Hazra2014ApJ,Albert2021}. The Babcock–Leighton mechanism for poloidal field generation in dynamo models of the solar cycle appear to be quite successful in explaining various manifestations of of solar magnetism \citep{Nandy2002,Jouve2008AA,Dasi2010A,Nandy2011,Cameron2015,Charbonneau2020}. Solar dynamo models can explain these different modes of solar activity fluctuation \citep{Tripathi2021} and are used widely in solar cycle prediction studies \citep{Bhowmik2018,Nandy2021SoPh}. Magnetic fields generated in the solar interior via the solar dynamo mechanism emerge on the surface and govern the structuring and evolution of the corona. The distribution of the large-scale coronal magnetic fields that expand into the heliosphere facilitates the flow of solar wind plasma into the inter-planetary medium and impacts the propagation of cosmic ray particles. Unlike sunspots, it is difficult to observe solar coronal magnetic fields directly due to low coronal plasma density and bright photospheric background. Hence we rely on data-driven modelling approaches to constrain the coronal fields. The evolution of OSF derived from coronal magnetic fields is an indicator of the solar forcing on the heliosphere. Past studies by \citet{Hoeksema1983JGR,Lee2011SoPh,Schatten1969,Schrijver2003,Yeates2010JGRA} show the evolution of the OSF over solar cycle time scales. 

Potential field source surface extrapolation (PFSS) is one of the widely used approaches to model the global coronal magnetic fields using photospheric magnetic field distribution as an input \citep{Altschuler1969,Schatten1969,Schrijver2003}. Assuming a vanishing current density, the PFSS model provides a unique magnetic field solution in a region between the surface at $r = R_{\odot}$ and a spherically symmetric source surface $r = R_{SS}$ where the magnetic fields are assumed to become purely radial. The unsigned magnetic flux through the source surface ($R_{SS}$) is defined as the OSF. The source surface height is typically set to be $R_{SS} = 2.5 R_{\odot}$ based on a comparison of PFSS models and observed coronal structures \citep{Hoeksema1983JGR,Schatten1969}. However, the source surface is believed to vary with solar magnetic activity levels. During solar minimum phases, the source surface moves closer to the surface \citep{Lee2011SoPh}. There exist multiple studies, wherein, the height of the source surface is adjusted in order to
minimize the misfit between the modeled and observed open solar flux at different phases of solar activity cycle. Earlier studies show a source surface height ranging from 1.8 - 2.5 $R_{\odot}$ provides a reasonable agreement between the modeled and observed open solar flux \citep{Lee2011SoPh,Arden2014JGRA,Badman2020ApJS,Reville2020ApJS}. Comparison between the PFSS-generated magnetic field and the observations of the interplanetary magnetic field provides an idea of the possible source surface height. Predictive approaches with coupled surface flux transport model and the PFSS extrapolation have shown great potential in reasonably accurate prediction of large-scale coronal magnetic field configurations \citep{Nandy2018,Dash2020}. In order to understand the coronal structure during a grand minimum phase, \citet{Riley2015} modeled coronal magnetic fields using global magnetohydrodynamic simulations. Global solar coronal configuration for past eclipses during grand minimum episodes is also explored by \citet{Lockwood2021A&G,Hayakawa2021,Hayakawa2020} using historic paintings.

Coronal magnetic field structure and its evolution modulate the state of the heliosphere which is reflected in the cosmic ray flux on earth. The modulation of cosmic rays due to solar activity is parameterized by the force-field approximation via the so-called modulation potential \citep[e.g,][]{Usoskin2005}. Using cosmogenic-isotope $^{14}$C measurements found in tree rings, the solar forcing on the cosmic rays is measured in terms of OSF and reconstructed sunspot number time series. \citet{Usoskin2021} reconstructed the OSF for the past $\sim$1000 years with annual cadence utilizing the cosmic ray flux assessed from cosmogenic-isotope $^{14}$C measurements in tree rings \citep{brehm21}. Apart from cosmic ray flux, in-situ magnetic field observations are also used to infer the evolution of OSF for past solar cycles \citep{Lockwood2014}. Long-term variations in OSF explain the variations in the state of the heliosphere due to solar forcing of coronal magnetic fields \citep{Krivova2021AA,Owens2013LRSP}. It is important to note that OSF is a spatially averaged quantity and hence lacks information about the large-scale coronal magnetic field structuring. Therefore modelling of coronal structures can provide critical insights to explain the solar forcing on these globally averaged quantities e.g. OSF and cosmic ray modulation potential.

In this study, we explore the coronal magnetic field configuration during regular activity, grand maximum and grand minimum phases. Our aim is to explore causal connections between the reconstructed solar activity and different phases of the solar dynamo by coupling the solar dynamo simulations to a PFSS model. We demonstrate differences in the nature of the coronal magnetic field configuration during these phases and discuss the resulting impacts on heliospheric parameters. Numerical model setups are explained in Section 2. We present our results explaining the variations in magnetic field strength, coronal magnetic field configuration, OSF and cosmic ray modulation potential for different solar activity episodes in Section 3. Finally, we conclude with a summary in Section 4.

\section{Theoretical modelling}

\subsection{Solar Dynamo model}
We simulate the Babcock-Leighton (BL) dynamo model \citep{Babcock1961ApJ,Leighton1969ApJ} with imposed stochastic fluctuations on the poloidal source. We utilize the 2.5D axisymmetric solar dynamo model -- \texttt{SURYA} which works in a kinematic regime \citep{Nandy2001ApJ, Nandy2002, Chatterjee2004}. The global magnetic field of the Sun can be expressed as a combination of poloidal and toroidal components in spherical polar coordinates i.e. 
\begin{equation}
    \mathbf{B} = B(r,\theta, t)\mathbf{e_{\phi}} + \mathbf{B_p},
\end{equation}
where $\mathbf{B_p} = \nabla \times [A(r,\theta, t)\mathbf{e_{\phi}}]$ corresponds to the poloidal component and $B (r,\theta, t)\mathbf{e_{\phi}}$ denotes the toroidal component, at any time instant $t$. The dynamo equations are obtained by solving the magnetic induction equation for $\mathbf{B}$ which can be written as,
\begin{eqnarray}
    \frac{\partial A}{\partial t} + \frac{1}{s}\Big[\mathbf{v_p}\cdot\nabla(s A)\Big] &=& \eta_p
    \left( \nabla^2 - \frac{1}{s^2}\right)A \nonumber \\ 
    &+& (\alpha_{MF} + \alpha_{BL}) B\,\,\,, \label{eq:1}\\
    \frac{\partial B}{\partial t} + s\left[\mathbf{v_p}\cdot\nabla\left(\frac{B}{s}\right)\right]
    + (\nabla\cdot\mathbf{v_p})B &=& \eta_t
    \left( \nabla^2 - \frac{1}{s^2}\right)B \label{eq:2} \\ 
&+& s (\mathbf{B_p}\cdot\nabla\Omega) + \frac{1}{s}\frac{\partial (sB)}{\partial r}\frac{\partial \eta_t}{\partial r}\,\,\,,\nonumber
\end{eqnarray}
where, $s=r \sin(\theta)$. Equation-\ref{eq:1} and \ref{eq:2} are solved on a uniform 129 $\times$ 129 grid between $0.55 R_{\odot} < r < R_{\odot}$ and $0 < \theta < \pi$. The poloidal field source term imbibes both the Babcock–Leighton mechanism ($\alpha_{BL}$) and mean-field $\alpha$-effect ($\alpha_{MF}$).

The meridional circulation which advects and distorts the magnetic field in each hemisphere is modeled through a single cell flow represented by $\mathbf{v_p}$  and  $\Omega (r,\theta)$ denotes the differential rotation in the solar convection zone (SCZ). In our model, we assume different magnetic diffusivities for the poloidal and toroidal field components, namely $\eta_p$ and $\eta_t$ respectively. The SURYA dynamo code uses different diffusivity terms for toroidal and poloidal field components to model the effect of suppression of turbulence by the relatively stronger toroidal component of the magnetic field \citep[see e.g.,][]{munoz-jaramillo2010}. The value for the $\eta_p$ and $\eta_t$ are set to be $2.6 \times 10^{12}$ cm$^{2}$s$^{-1}$ and $4.0 \times 10^{10}$ cm$^{2}$s$^{-1}$ respectively. Specifically, our model is adapted from \citet{Pasos2014} which introduced stochastic fluctuations and an additional mean-field $\alpha$-effect for recovery of cycles from grand minima episodes. The mean-field $\alpha$-effect is distributed through the bulk of the solar convection zone and is quenched by fields exceeding 10$^4$ G whereas the Babcock-Leighton $\alpha$-source ($\alpha_{BL}$) operates near the surface. For further details refer to \citet{Pasos2014}. The presence of both the Babcock-Leighton poloidal source and mean-field $\alpha$-effect add interesting aspects to the dynamics of the solar cycle as recently explored in the context of solar cycle predictability \citep{hazra2020}. 


The idea behind introducing stochasticity in the Babcock-Leighton $\alpha$-source ($\alpha_{BL}$) is to mimic the modulation in the surface BL mechanism by scatter (around the mean Joy's law) in tilt angle of emerged bipolar sunspot pairs. Likewise, fluctuations in the mean-field $\alpha$-source ($\alpha_{MF}$) dictate the impact of turbulent helical convection in the deep interior.  The combined action of the poloidal source terms facilitates recovery from grand minimum \citep{Pasos2014,Hazra2014ApJ}. The Babcock-Leighton $\alpha$-effect is defined by:
\begin{eqnarray}
    \alpha_{BL}(r,\theta) = \alpha_{BL}^{0}\frac{ \cos \theta }{4} \left[1+\textrm{erf}
    \left( \frac{r-r_1}{d_1}\right)\right] \times\ \left[1-\textrm{erf}
    \left( \frac{r-r_2}{d_2}\right)\right] \nonumber \\
   \times \, a_1\, \left[ 1 + \textrm{erf}
    \left(\frac{B_\phi^2 - B_{1lo}^2}{d_3^2}\right)\right]\times
    \left[ 1 - \textrm{erf}\left(\frac{B_\phi^2 - B_{1up}^2}{d_4^2}\right)\right]\,\,\,.
    \label{c6.eq-alphastandard}
\end{eqnarray}

Upper and lower quenching terms $B_{1up}$, $B_{1lo}$ are introduced in equation-\ref{c6.eq-alphastandard}. Such parametrization of the $\alpha_{BL}$ is important from a physical perspective. On the one hand, toroidal fields which are very weak do not coherently emerge to produce bipolar sunspot pairs and thus cannot contribute to poloidal field generation. On the other hand, very strong toroidal fields which do not pick up significant tilt do not produce significant polar fields. The values of the constants are set to $r_1 = 0.95 R_{\odot}, r_2 = R_{\odot}$ and $d_1 = d_2 = 0.025 R_{\odot}$. The $\alpha_{BL}^{0} = 27$m s$^{-1}$ controls the amplitude of the source term; $a_1 = 0.393$ is a normalization constant; the lower threshold $B_{1lo} = 10^3$ G; $d_3 = 10^2$ G; the upper threshold $B_{1up} = 5 \times 10^5$ G and $d_4 = 10^6$ G. The detail description of the quantities mentioned in equation-\ref{c6.eq-alphastandard} and their mathematical parametrization, are available in \citet{Pasos2014}.

The dispersion in the poloidal source term distribution controls the poloidal field amplitude and thus the solar cycle strength. Random stochastic fluctuations of strength 150\% are introduced around the mean value of the $\alpha_{BL}$, denoted by $\alpha_{BL}^0$, so that $\alpha_{BL}$ = $\alpha_{BL}^0$ + $\alpha_{BL}^{fluc}$ $\sigma$(t,$\tau$). Here, $\alpha_{BL}^0$ is set to $27$ m s$^{-1}$ and $\sigma$(t,$\tau$) is assigned random values between [-1.5, 1.5] after each coherence time $\tau$ (here 6 months). $\alpha_{BL}^{fluc}$ is set to the same value as $\alpha_{BL}^{0}$.

For the buoyancy algorithm, our model searches for toroidal field exceeding a critical threshold (${B_c} = 8 \times 10^4$ G) at the base of the convection zone (at $r = 0.71 R_{\odot}$). When this condition is satisfied, the algorithm removes half of the toroidal field and deposits this field near the surface where the near surface $\alpha$-effect acts locally on the toroidal field to mimic the Babcock-Leighton mechanism. The eruptions are proxies for sunspots in our simulation setup and their latitude and time of emergence are used to reconstruct the simulated butterfly diagrams \citep{Nandy2001ApJ, Nandy2002, Chatterjee2004}. 

The number of sunspot eruption proxies is counted using the aforementioned buoyancy algorithm to model the sunspot proxy time series. We also calculate the surface radial magnetic field using the magnetic vector potential A using the following expression:
\begin{eqnarray}
    {B_r} = \frac{ 1 }{R \sin\theta} \left[\frac{\partial (A\sin\theta)}{\partial \theta}\right] \,\,\,\,.
    \label{br}
\end{eqnarray}

We performed a 6000-year-long solar dynamo simulation. However, our dynamo simulation is not calibrated to observations and hence is not suitable for direct comparison with observed quantities. In this paper, we intend to understand the qualitative nature of the magnetic field evolution during grand maxima, grand minima and regular activity phases.

\subsection{Solar coronal magnetic field model}
Reconstruction of large scale coronal magnetic fields can be done following the Potential Field Source Surface extrapolation (PFSS) technique \citep{Schatten1969,Altschuler1969} utilizing the surface radial magnetic field as a lower boundary condition. This technique assumes the solar corona to be current-free till a spherical source surface (of radius $r={R}_{ss}$). Beyond the source surface, the impact of solar wind is dominant which makes the magnetic field lines purely radial. Hence in the region $R_{\odot} \leqslant r  \leqslant {R}_{\mathrm{ss}}$, 
\begin{eqnarray}
    \nabla \times \mathbf{B} &=& 0.
\end{eqnarray}
We can express the magnetic field in terms of a scalar potential $\phi$ that satisfies, 
\begin{eqnarray}
    -\nabla \phi &=& \mathbf{B}.
\end{eqnarray}
Since $\nabla \cdot \mathbf{B} = 0$ we can write,
\begin{eqnarray}
    \nabla^2 \phi &=& 0.
\end{eqnarray}
By solving for the scalar potential ($\phi$) we can compute the solar coronal magnetic fields within the source surface. We use the radial magnetic field at the surface computed by the solar dynamo model as the lower boundary condition of the PFSS model. This coronal magnetic field modelling technique is widely used in the solar physics community to compute the large scale configuration of the corona. For a detailed derivation of the model equations, refer to \citet{Schrijver2003}. OSF is calculated by integrating the unsigned radial magnetic field over the source surface.

 
\subsection{Cosmic ray modulation potential}
The cosmic ray modulation potential describes the mean deceleration (i.e. loss of energy/rigidity) of galactic cosmic-ray particles within the heliosphere modulated by solar activity. The process of the heliospheric cosmic-ray modulation is complex \citep{Potgieter2013} but is often parameterized via heliospheric parameters such as OSF and the cosmic ray modulation potential \citep[e.g.][]{Usoskin2002,Wu2018}. Solar forcing on the cosmic-ray modulation (parameterized as the heliospheric modulation potential) is mediated via the OSF. We calculate cosmic ray modulation potential ($\Phi$) using a semi-empirical formalism given by \citet{Asvestari2016},
\begin{eqnarray}
    \Phi = \Phi_0 \times F^{n-\frac{\theta^{'}}{\theta_0^{'}}} (1-\beta p).
	\label{eq:pot_asvestari}
\end{eqnarray}
Here $\Phi$ and $F$ denote the cosmic ray modulation potential and OSF respectively. The heliospheric tilt angle ($\theta^{'}$) denotes the average angle between the heliospheric current sheet and the equatorial plane. In our model, $\theta^{'}$ is the angle subtended by the line joining the Sun's center and the source surface neutral line (i.e. where $B_r = 0$ G) with respect to the solar equator. The free parameters $\Phi_0, \theta_0^{'}, n$ and $\beta$ are adopted from \citet{Asvestari2016} as: $\Phi_0 = 1473.9$ MV, $\theta_0^{'} = 150^{\circ}$, $n = 1.03$ and $\beta = 0.095$. The polarity of the solar magnetic field ($p$) is assigned a value $p$= +1 (positive)/ -1 (negative) depending on the sign of the polar fields (for our calculations we consider the sign of the solar north pole as reference). The OSF ($F$) is calculated at the source surface $R_{ss} = 2.5 R_{\odot}$ and normalized with the maximum value. This normalized OSF is used to compute the cosmic ray modulation potential.

In our 6000-year simulation the normalization factor for the OSF is $1.8 \times 10^{24}$ Mx. Using the OSF, we compute the cosmic ray modulation potential for 6000-year dynamo simulation.

\section{Results}
In our long-term solar dynamo simulation, we find random occurrences of fluctuating solar activity phases -- ranging from grand maxima to grand minima. Solar activity fluctuations can be broadly divided into two classes namely grand minima phase and regular activity phase \citep{Usoskin2016AA,Wu2018}. The regular activity phase includes grand maximum like an enhanced magnetic activity period and the moderate activity phases. Likewise, the periods where the solar magnetic activity level is minimum are categorised as grand minima phases.

\begin{figure*}
	\centering{\includegraphics[width=0.9\textwidth]{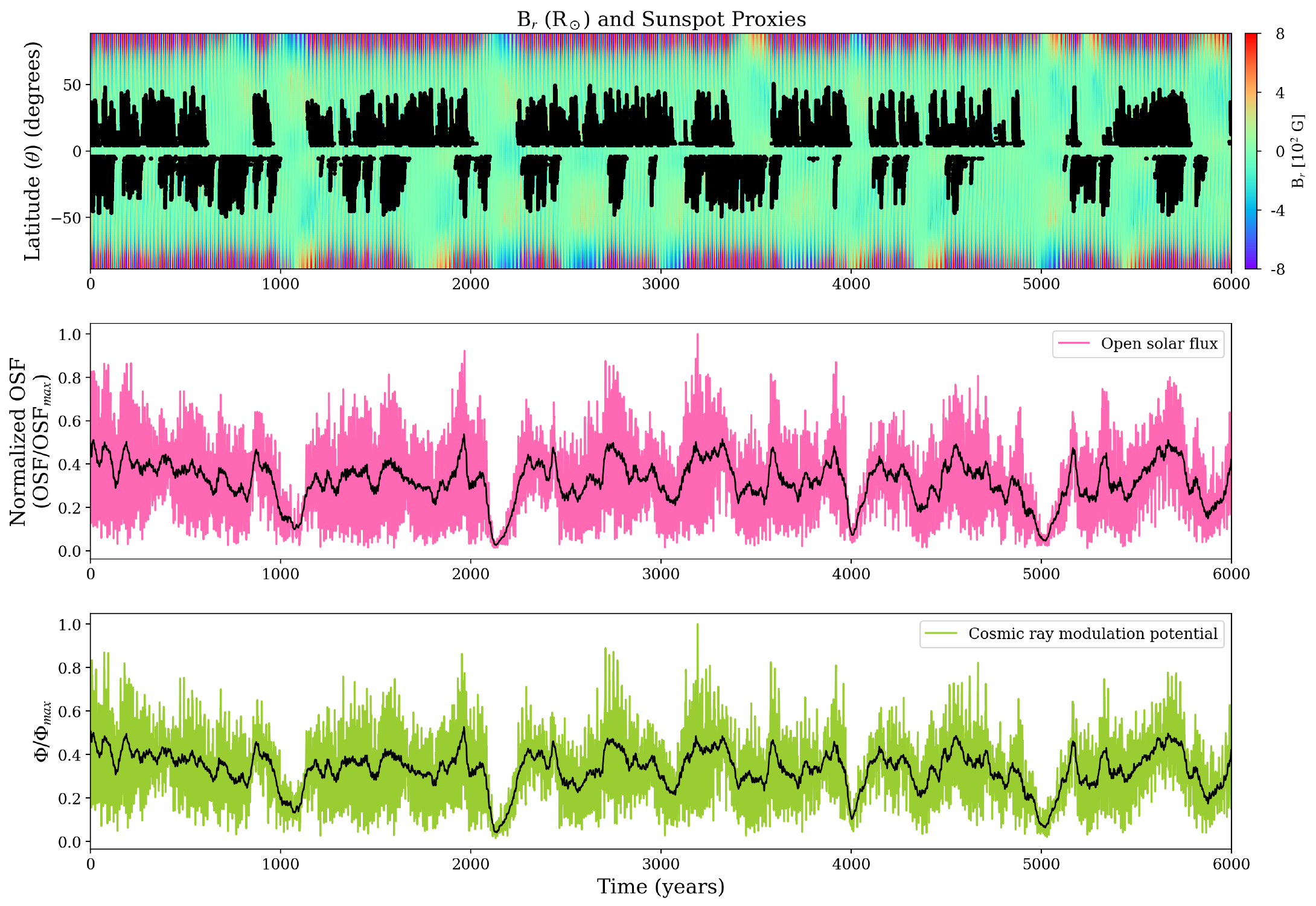}}
    \caption{Long-term stochastically forced solar dynamo simulation for 6000 years. The top panel shows a butterfly diagram of the surface radial magnetic field ($B_r$). The emergence latitudes of the sunspot eruption proxies are overplotted in black. The middle panel denotes normalized open solar flux computed using solar dynamo simulation and PFSS model in the magenta curve. A 22-year running average of the modeled OSF time series is plotted in the solid black curve. In the lower panel, we plot the modeled normalized cosmic ray modulation potential in solid green color and the corresponding 22-year running average in solid black. For the grand minimum phase, sunspot eruption proxies are absent on the solar surface. Hence, the resulting heliospheric modulation due to solar activity variation which is indicated by the OSF and the cosmic ray modulation potential shows a drop in magnitude for these phases. Similarly, we find signatures of enhanced OSF and cosmic ray modulation potential corresponding to grand maximum phases (periods of higher solar activity) in our modeled output.}
    \label{fig:1}
\end{figure*}

The evolution of radial magnetic field on the solar surface ($B_r(R_{\odot})$) is shown in the butterfly diagram in the top panel of Fig. \ref{fig:1}. The overplotted sunspot eruption proxies (plotted in black circles) denote the level of solar activity variation. In the butterfly diagram, the periods devoid of sunspot eruption proxies correspond to grand minima episodes. We find four such grand minima phases from Year 998  to 1139, Year 2099 to 2244, Year 3395 to 4098 and Year 4904 to 5119, where sunspot eruption proxies are absent in both the hemispheres. For such phases, the heliospheric modulation parameters e.g. OSF and the cosmic ray modulation potential show a dip (middle and bottom panel of Fig. \ref{fig:1}). Additionally, there are several other instances of hemispheric grand minimum where sunspot eruption proxies are absent in either one of the hemispheres.

\begin{figure*}
	\includegraphics[width=\textwidth]{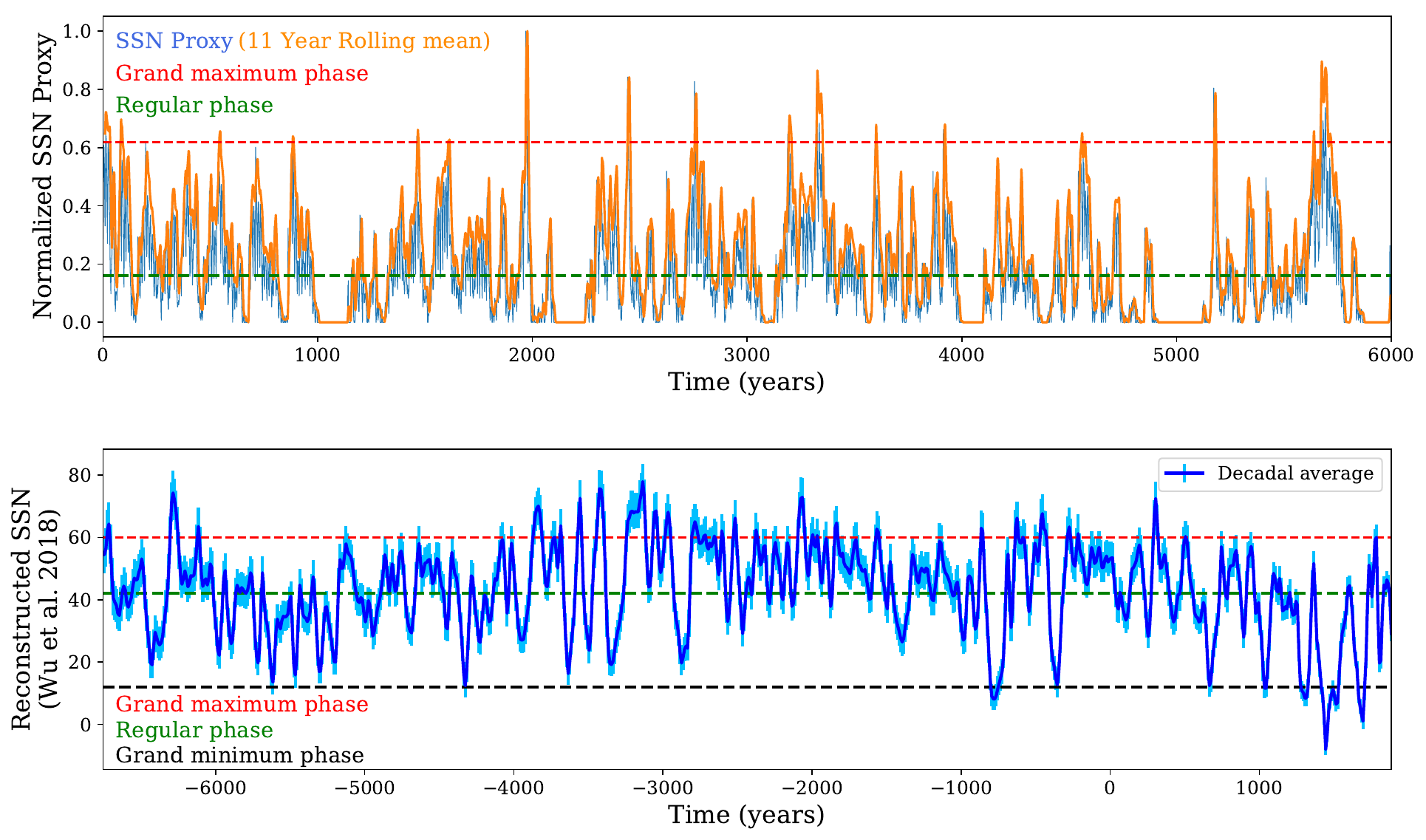}
    \caption{Sunspot number time series. Normalized time series of solar dynamo generated sunspot eruption proxy for 6000 years is shown in blue in the top panel. The eleven-year running average of the SSN is plotted in orange. The mean sunspot number is plotted in the green dashed curve. Episodes where the number of sunspots is greater than mean+3$\sigma$ (shown in the red dashed curve), are identified as grand maxima. Phases with no sunspot eruption proxies are the grand minimum phases in our simulation. In the bottom panel reconstructed decadal averaged SSN \citep{Wu2018} is shown in the blue curve. Here the minimum threshold for the grand maximum phases is denoted by the red dashed curve. The green dashed curve shows the main component (normal/moderate phase) and the black dashed line denotes the grand minimum component of the reconstructed solar activity cycle. There are multiple grand maxima and grand minima phases present in both reconstructed time series and our long term solar dynamo simulation.}
    \label{fig:2b}
\end{figure*}

In order to identify grand minimum, regular activity and grand maximum phase in our model output, we analyse the time series of sunspot eruption proxies. The mean value of the simulated SSN proxy time series is shown in the green dashed line in the top panel of Fig. \ref{fig:2b}. We define a threshold of mean sunspot number + 3$\sigma$ (red dashed line in Fig. \ref{fig:2b}) to identify the grand maximum phases.  For grand minimum phases, we identify the epochs in the time series where there are no sunspot eruptions (at least for three consecutive cycles). Lastly, for the regular activity phases, the sunspot eruption proxy remains close to the global mean of the sunspot proxy (green dashed line in Fig. \ref{fig:2b}). With this definition, we find multiple instances of grand maxima, grand minima and regular activity phases in our simulation. 

In the long term reconstruction of solar activity cycle over nine millennia, there are multiple occurrences of such fluctuating solar activity phases \citep{Wu2018}. This reconstruction provides decadal averaged reconstructed SSN (see bottom panel of Fig. \ref{fig:2b}). Different components of the solar activity i.e. grand maximum, grand minimum, normal (moderate) phase are identified in red, black and green dashed curves respectively in the bottom panel of Fig. \ref{fig:2b}. We analyse these different phases in terms of coronal magnetic field configuration.

\subsection{Solar corona during grand maxima, grand minima and a regular solar activity phase}
\begin{figure*}
	\centering{\includegraphics[width=0.8\textwidth]{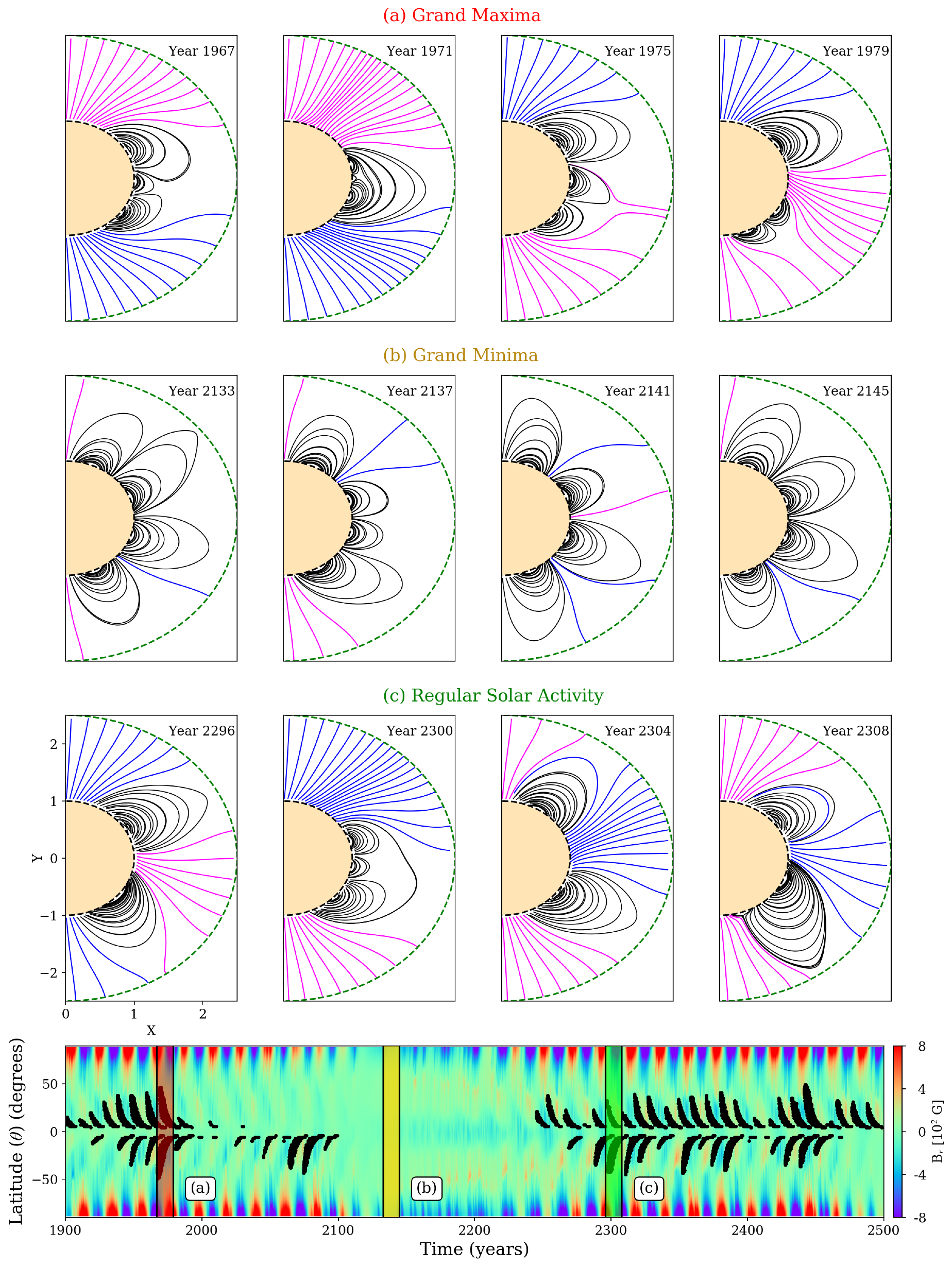}}
    \caption{Evolution of solar coronal magnetic field configuration during grand maxima, grand minima and regular solar activity phase. In the bottom panel, the surface radial magnetic field butterfly diagram is shown and the sunspot eruption proxies are plotted in black color. Shaded regions on the butterfly diagram denote the grand maxima (red), grand minima (yellow) and regular solar activity (green) phase in the long-term solar dynamo simulation. They are labelled as (a), (b) and (c) respectively. For each of these phases, we show the corresponding coronal magnetic field configuration starting from the cycle minimum (T0) in four increments e.g. for the grand minimum episode distribution of coronal magnetic field for the year 2133, 2137, 2141, 2145 is shown where year 2133 and 2145 correspond to cycle minimum. Extrapolated coronal magnetic field lines are shown with open field lines denoted by blue (radially outward) and magenta (radially inward) curves. Closed magnetic field lines are plotted in black. A complex coronal configuration consisting of closed magnetic field lines reaching close to high-latitude regions is observed during the modeled grand minimum phase. We have fixed the source surface height at $r = 2.5 R_{\odot}$ for all the phases to maintain consistency. Source surface ($r = 2.5 R_{\odot}$) is denoted in the green dashed line for all the cases. Detailed analysis of the coronal magnetic field configuration for different epochs is provided in the text.}
    \label{fig:2}
\end{figure*}

The modulation of the state of the heliosphere by the solar coronal magnetic fields can be parameterized by OSF which in turn impacts the cosmic ray modulation. OSF and the cosmic ray modulation potential are spatially averaged quantities. Hence they lack a description of solar coronal magnetic fields (whether they are open or closed within the solar source surface) that govern the flow of the solar wind in the heliosphere. In order to understand the spatial distribution of the fields during different solar activity phases, we choose a grand maximum (from Year 1967 to 1979), grand minimum (from Year 2133 to 2145) and a regular activity phase (from Year 2296 to 2308) for further analysis. This choice of the solar activity phases is consistent with our definition based on the modeled sunspot eruption proxy time series. The coronal magnetic field configuration obtained using the PFSS model is provided in Fig. \ref{fig:2} for corresponding years. Grand maxima, grand minima and regular activity phase are denoted as shaded regions on the simulated $B_r (R_{\odot})$ butterfly diagram in the bottom panel as (a), (b) and (c) respectively. For each of these episodes, coronal magnetic field configuration is shown at a four-year time interval starting from the corresponding cycle minimum. We fix the source surface height at $r = 2.5 R_{\odot}$ for consistency in our long term simulation.

Near the cycle minimum in a grand maximum phase, we find the global corona to be in a dipolar state. We also find that the high latitude regions are densely populated with open field lines (magenta for radially inward and blue for radially outward field lines) whereas the closed field lines are localised near the equatorial region (black for closed field lines) as shown in the panel (a) Grand Maxima of Fig. \ref{fig:2}. For this phase, the global configuration of the corona shows a combination of closed and open field lines. 

During the chosen grand minimum phase in our simulation the sunspot eruption proxies are absent in both the hemispheres indicating a reduced solar activity. The overall strength of radial magnetic field is also significantly low as compared to other phases which will be discussed later. The resulting coronal magnetic field configuration as depicted in Fig. \ref{fig:2} (b) shows highly populated closed field regions in our model. We find the presence of these closed magnetic fields reaching high latitudes persisting throughout the grand minimum phase. Note that the variation of the source surface height during grand maximum and minimum can impact the coronal magnetic field configuration as discussed in Section \ref{sec:3.4}.

For a regular solar activity phase (Fig. \ref{fig:2} (c)) we observe the same polarity open field lines near both the poles at cycle minimum for Year 2296. This demonstrates signatures of the quadrupolar parity in coronal magnetic fields. The global corona shifts to a dipolar parity after four years (Year 2300). In our solar dynamo model, we incorporate fluctuations in the poloidal source term for the Northern and Southern hemispheres independently. A combination of hemispheric decoupling due to stochastic fluctuations and non-linear terms in our model possibly contributes to parity modulation (see \citet{Hazra2019} for an extensive study focusing on this aspect). Independent studies by \citet{Knobloch1998MNRAS,Beer1998SoPh} also find parity modulation due to co-existing interacting dynamo modes. During grand minima, regular activity and grand maxima phase the modulations in the coronal magnetic field configuration are caused by the variation in the surface magnetic field configuration.

\begin{figure*}
	\includegraphics[width=\textwidth]{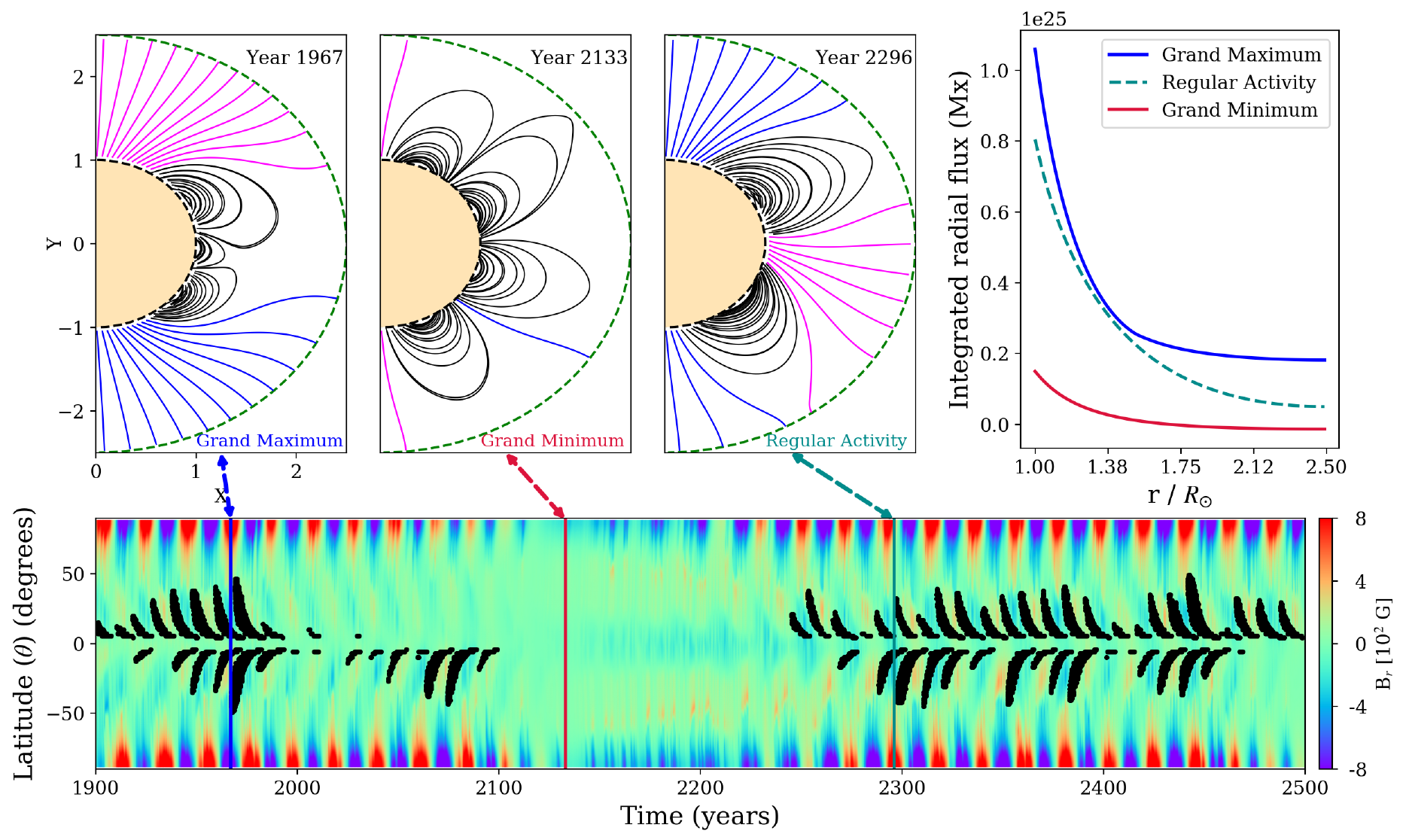}
    \caption{Reduction in magnetic field strength for different phases of solar activity. In our analysis year 1967 corresponds to a grand maxima phase. Similarly, year 2133 and 2296 denote grand minimum and regular solar activity phase. The integrated unsigned radial flux ($\int |B_r| 2\pi r^2 \sin \theta d\theta$) across different radial heights are plotted in the top-right panel in blue (grand maxima), dashed dark cyan (regular activity) and red (grand minimum). The decrease of over all flux from grand maxima to grand minima is observed for these three cases. In coronal magnetic field configuration plots source surface ($r = 2.5 R_{\odot}$) is denoted by green dashed line. Distribution of closed and open magnetic fields for different phases (grand maximum, grand minimum and regular) is provided with radially outward/inward open field lines plotted in blue/magenta. On the butterfly diagram (in the bottom panel) these phases are marked in solid lines for ease of understanding.}
    \label{fig:2c}
\end{figure*}

The surface magnetic field distribution shapes the coronal magnetic fields which in turn modulates the heliospheric parameters (e.g. OSF and cosmic ray modulation potential). In order to understand the variation of solar magnetic activity we plot the variation of integrated unsigned radial flux with height for grand maxima (Year 1967), grand minima (Year 2133) and regular activity phase (Year 2296) in the top-right panel of Fig. \ref{fig:2c}. The integrated unsigned radial flux is computed as $\int |B_r (r,\theta)| 2 \pi r^2 \sin \theta d\theta$ for different heights. Among these three different phases, the integrated unsigned radial flux is highest for the grand maximum and lowest during the grand minimum. The integrated unsigned radial flux for the regular phase lies between the grand maximum and grand minimum.

\subsection{Solar coronal magnetic field configuration during cycle maximum and minimum phases}
\begin{figure*}
	\centering{\includegraphics[width=0.9\textwidth]{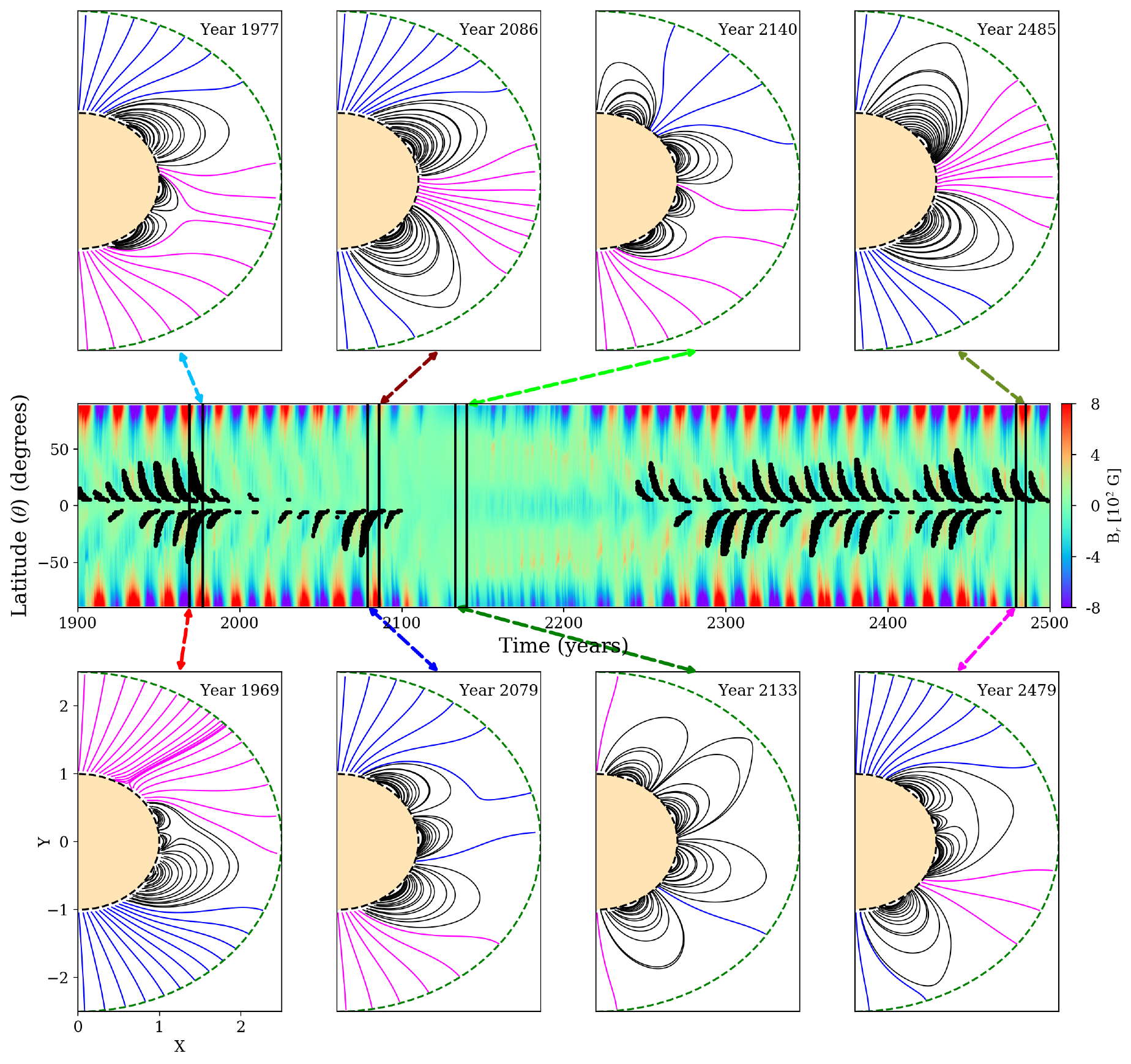}}
    \caption{Solar coronal magnetic field configuration for different phases based on hemispheric sunspot eruptions. The butterfly diagram captures a time frame spanning over 600 years starting from year 1900 until year 2500 with sunspot eruption proxies over-plotted in black from SURYA 2.5D kinematic solar dynamo model. Global solar coronal magnetic field configuration is plotted for different cases. The black curve denotes closed magnetic fields and the magenta/blue lines show radially inward/outward open field lines. Year 1969 and 1977 demonstrate a period where the sunspot eruptions are observed till high latitudes in both hemispheres in our simulation. For this case the coronal magnetic field distribution indicates the presence of complex coronal loops closer to the equator. Year 2079 and 2086 denotes a period where sunspot eruption proxies are absent only in the northern hemisphere. The resulting coronal magnetic field distribution shows a shift in parity (as the global dipolar structure is changed to a quadrupolar configuration in year 2086). For Year 2133 and Year 2140 sunspot eruption proxies are absent in both hemispheres. For this case, the extrapolated magnetic field distribution is quite complex. We notice closed magnetic field lines near polar high latitudes. We have assumed a fixed source surface height of $r = 2.5 R_{\odot}$ for our modelling irrespective of the solar activity phase. Similarly, for a case where sunspot eruption proxies are absent in the southern hemisphere (Year 2479 and 2485), we observe complex coronal magnetic field structure along with a shift in parity in large scale solar magnetic field. The parity modulations result from the hemispheric decoupling and non-linear nature of the solar dynamo mechanism.}
    \label{fig:3}
\end{figure*}

In Fig. \ref{fig:3}, the coronal magnetic field configuration for different cases (based on the hemispheric appearance of sunspot eruption proxies) are presented at the corresponding cycle maximum and minimum. We choose a period from Year 1900 to 2500 for our analysis. Sunspot eruption proxies are denoted in dark circles over the butterfly diagram of surface radial magnetic field i.e. $B_r (R_{\odot})$ in the middle panel of Fig. \ref{fig:3}. For a phase, where the numbers of sunspot eruptions are large in both hemispheres (close to a grand maximum phase), we find the eruptions reach a higher latitude compared to other phases. In such a phase (Year 1969), the coronal magnetic field configuration also shows a combination of closed and open field lines. The polarity of the polar fields is reversed at the end of the cycle (Year 1977) which can be seen in the polar coronal field lines. For a period where sunspot eruptions are absent only in the northern hemisphere (Year 2079 and Year 2086), we find a shift in parity -- from dipolar to quadrupolar --  within the cycle. When the sunspot proxies are absent in both the hemispheres (Year 2133 and Year 2140) the global magnetic field configuration becomes complex. We find the presence of more closed fields near high-latitude regions for such phases. When the eruptions cease in the southern hemisphere only (Year 2479 and Year 2485), we find the signature of parity modulation in the coronal magnetic fields. Such parity reversals in long term solar dynamo simulations are reported by \citet{Hazra2019}. Origin and impact of asynchronous solar activity across hemispheres have been also explored independently by \citet{Kitchatinov2021ApJ,Shukuya2017ApJ,Suchssler2018AA,Norton2014SSRv}.

\subsection{Variation of source surface height during grand minima and grand maxima phases}
\label{sec:3.4}
The outer boundary of the PFSS model i.e. the source surface is a spherical surface located at a constant height from the solar surface. At the source surface, the magnetic field is assumed to be purely radial and within the source surface, the coronal field is current free. In most studies, the value of source surface height ($r = 2.5 R_{\odot}$) is fixed to match the observed interplanetary magnetic field pattern \citep{Hoeksema1983JGR}. However, \citet{Schatten1969,Levine1982SoPh,Levine1977ApJ,Lee2011SoPh} found that for low solar activity periods lowering the source surface height better matches the observations of open magnetic fields. Since our simulation covers periods of both grand maximum and minimum, we vary the source surface height in our simulation and explore the resulting changes in the coronal magnetic field configuration. Nevertheless, for the long-term computations source surface is fixed at $R_{SS} = 2.5 R_{\odot}$.

\begin{figure*}
	\centering{\includegraphics[width=0.9\textwidth]{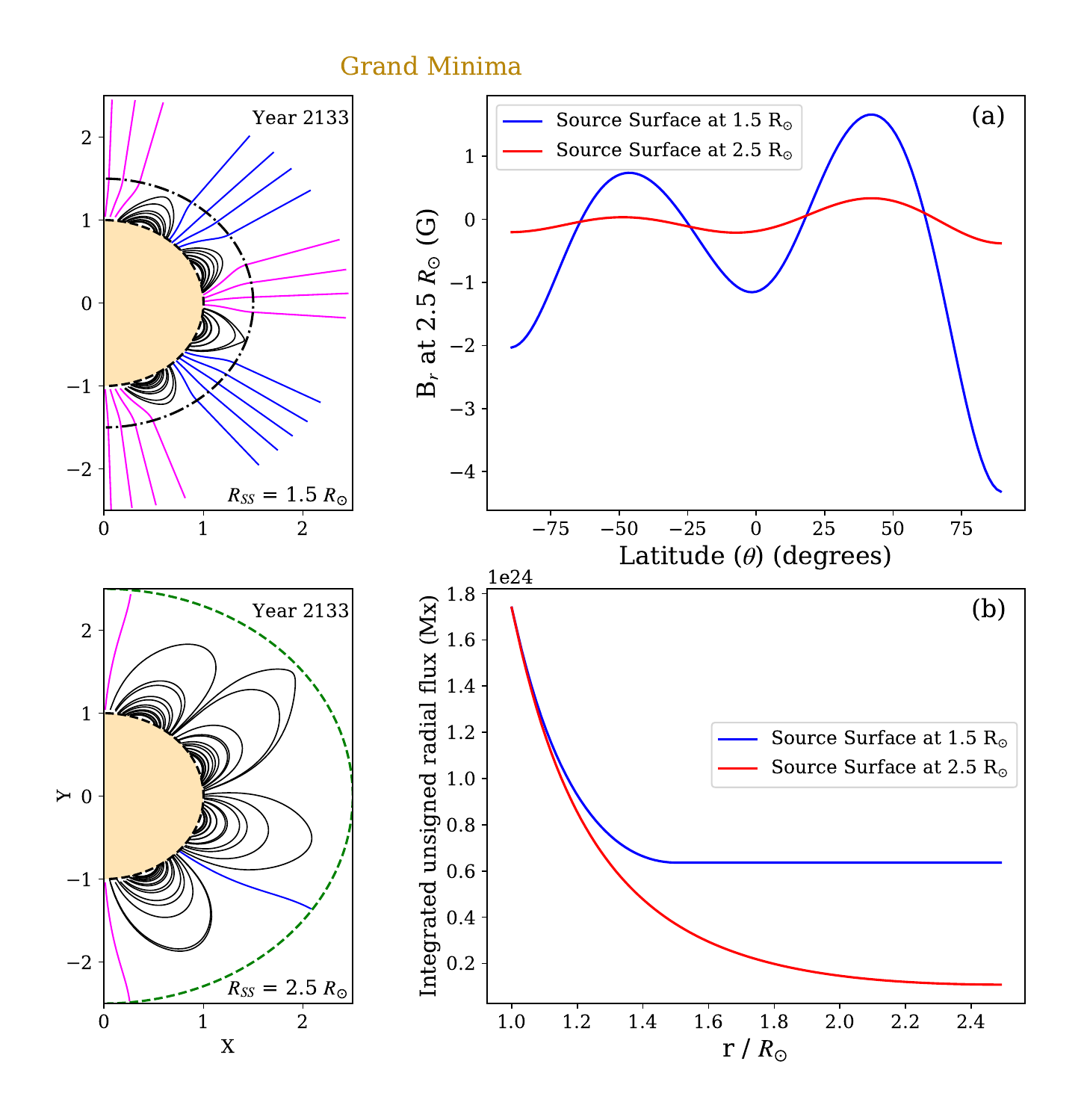}}
    \caption{Coronal magnetic field distribution during a grand minima episode (year 2133) for two different source surface heights $r = 2.5 R_{\odot}$ and $1.5 R_{\odot}$. Green dashed curve and black dashed-dotted curve in the bottom panel identify the source surface height at $r = 2.5 R_{\odot}$ and $r = 1.5 R_{\odot}$ in our model. (a) Extrapolated radial field ($B_r$) at $r = 2.5 R_{\odot}$ as a function of latitude for $R_{SS} = 1.5 R_{\odot}$ plotted in blue and for $R_{SS} = 2.5 R_{\odot}$ plotted in red. (b) Variation of integrated unsigned radial flux at different heights when $R_{SS}$ is at $1.5 R_{\odot}$ (in blue) and $1.5 R_{\odot}$ (in red). By lowering the source surface closer to the surface we observe more open field lines in the modeled output. However, the closed magnetic field lines still persist within the spherical source surface that may not have been captured in the historical total solar eclipse paintings. Source surface height is one of the key parameters for the calculation of OSF. Choosing a source surface distant from the surface during grand minimum phases results in a lower OSF value.}
    \label{fig:3a}
\end{figure*}

\begin{figure*}
	\centering{\includegraphics[width=0.9\textwidth]{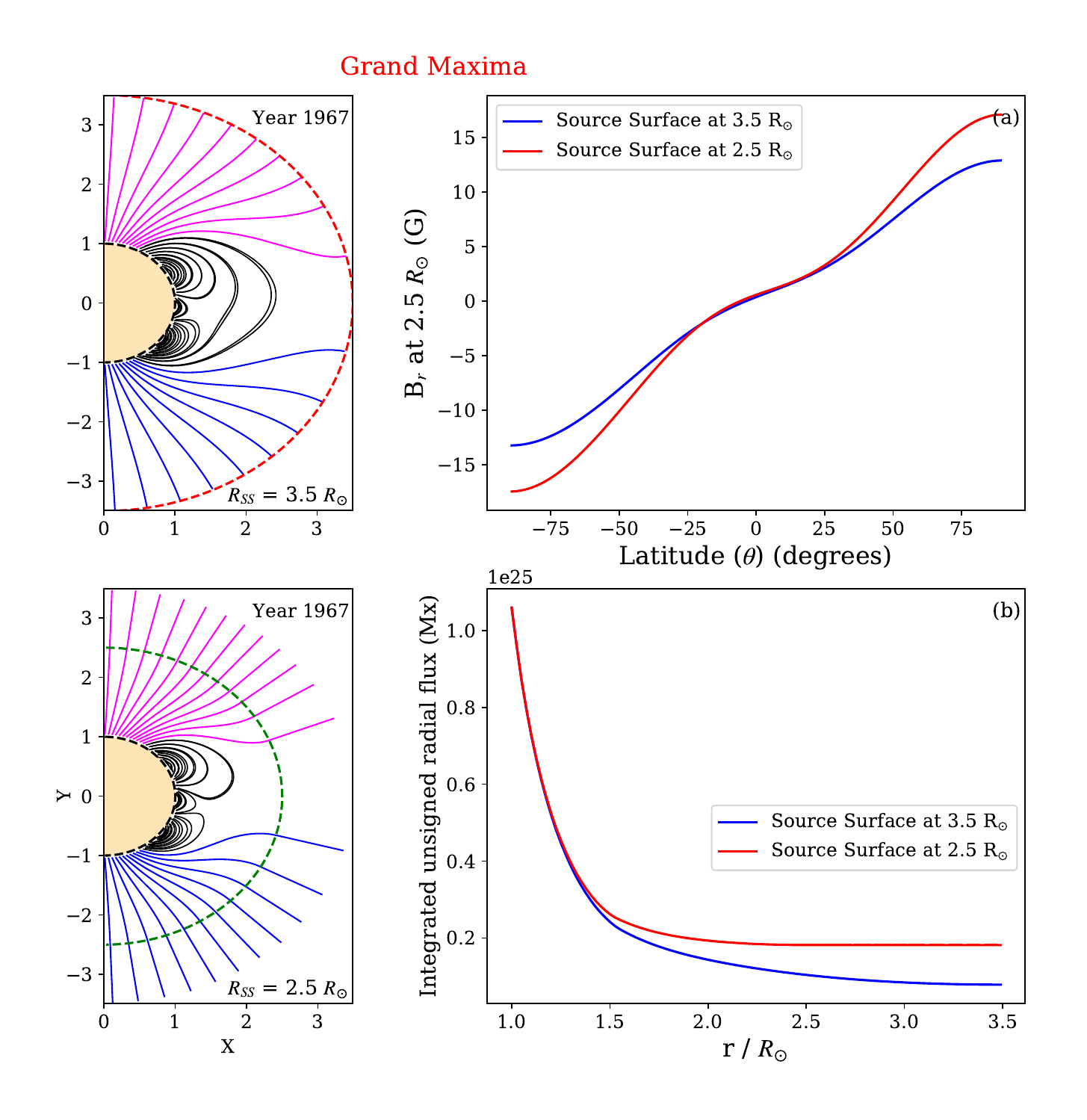}}
    \caption{Distribution of magnetic field lines during a grand maxima episode (year 1967) for two different source surface heights $r = 3.5 R_{\odot}$ and $2.5 R_{\odot}$. The green/red dashed curves in the plots show the source surface height at $r = 2.5 R_{\odot}$ and $r = 3.5 R_{\odot}$. (a) Modeled radial field at $r = 2.5 R_{\odot}$ as a function of latitude for $R_{SS} = 3.5 R_{\odot}$ plotted in blue and for $R_{SS} = 2.5 R_{\odot}$ plotted in red. (b) The integrated radial at different heights when $R_{SS}$ is at $3.5 R_{\odot}$ (in blue) and $2.5 R_{\odot}$ (in red). Lowering of source surface height during a grand maximum phase results in the opening up of more field lines which results in a higher radial field near the outer boundary and OSF value.}
    \label{fig:3b}
\end{figure*}

For the grand minimum phase, we do some additional extrapolation by setting the source surface height at $r = 1.5 R_{\odot}$ and $r = 2.5 R_{\odot}$ in Fig. \ref{fig:3a}. The source surface heights are marked in the black dashed-dotted curve for $r = 1.5 R_{\odot}$ and in the green dashed curve $r = 2.5 R_{\odot}$. We find the presence of open magnetic fields for a lower source surface height. However, the low-lying closed magnetic fields are persistent in this case as well. We also show the extrapolated radial magnetic fields at $r = 2.5 R_{\odot}$ for both cases (see Fig. \ref{fig:3a} (a)). It is to note that varying the source surface height which is an un-constrained parameter (set to 2.5 R$_{\odot}$ for the whole simulation duration) may lead to a change in magnetic fieldline connectivity at a global coronal scale. Lowering the source surface height results in a higher radial field value during a grand minimum phase. For the integrated unsigned radial flux we find a similar trend as well (see Fig. \ref{fig:3a} (b)). Similarly, for a grand maximum phase, we extrapolate the coronal magnetic fields for two different source surface heights $r = 3.5 R_{\odot}$ and $r = 2.5 R_{\odot}$ (see Fig. \ref{fig:3b}). The source surface heights are shown in red and green dashed curves here. Upon calculating the radial magnetic field at $r = 2.5 R_{\odot}$ (see Fig. \ref{fig:3b} (a)) and the integrated unsigned radial flux (see Fig. \ref{fig:3b} (b)) we find a relatively higher flux near the boundary. In our long term simulation, solar activity variations cover a wide range of activity episodes. Hence for PFSS modelling, fixing the source surface height at $r = 2.5 R_{\odot}$ can result in a higher OSF during the grand maximum phase and lower OSF for the grand minimum phase.

\subsection{Suppressed 11-year cycle period during grand minimum episodes}
\begin{figure*}
	\includegraphics[width=0.9\textwidth]{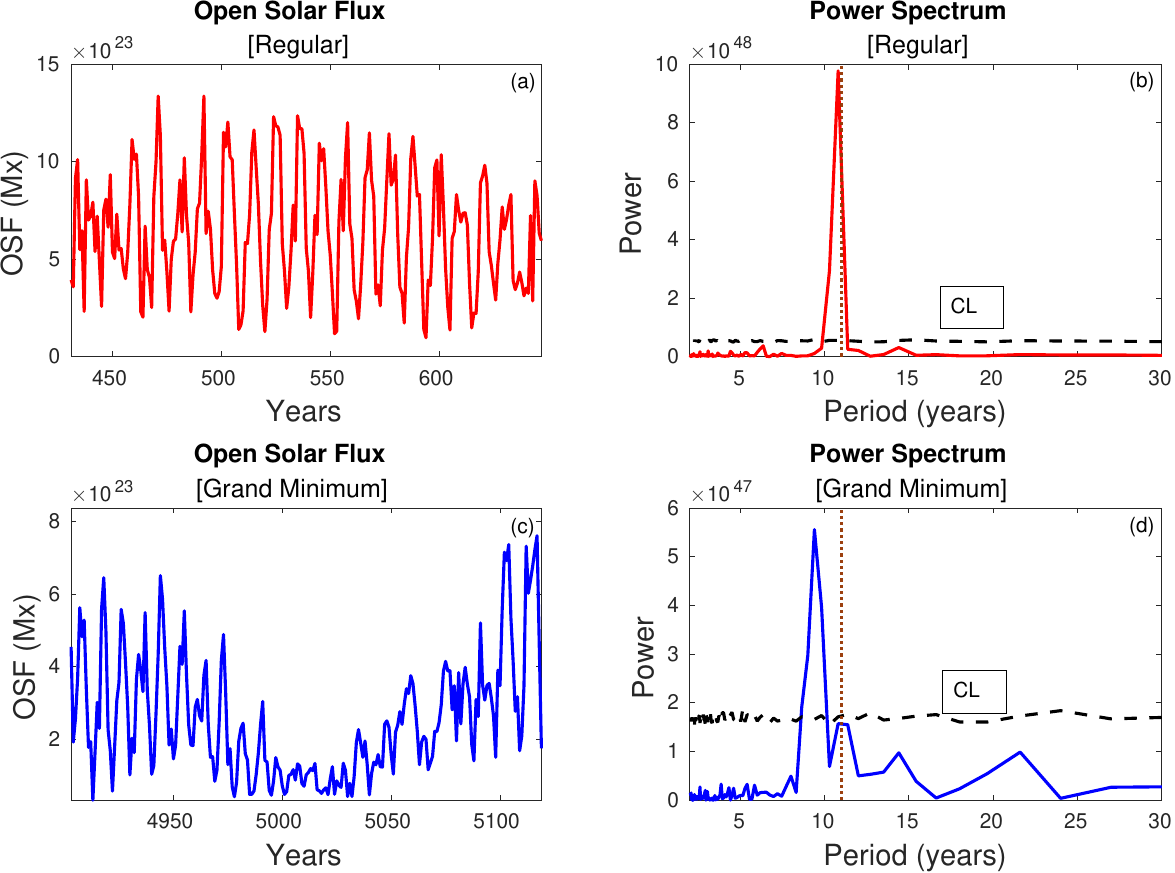}
    \caption{Spectral analysis on the simulated OSF time series. (a) The red curve indicates the time series of OSF for a regular activity phase (where there are finite sunspot eruption proxies on the solar surface) from year 432 to 648. (b) The red curve shows the power spectrum of OSF for the regular activity phase. FFT window size was chosen to be 1-year for our calculations. The overplotted black dashed line denotes the upper 95th percentile of the FFT spectra for 1000 re-sampled OSF time series for the regular activity phase. (c) The blue curve shows the simulated OSF time series for one of the grand minimum episodes (year 4904 to 5120). (d) The power spectrum of OSF for the grand minimum phase. The black dashed line here depicts the upper 95th percentile of the FFT spectra for 1000 re-sampled OSF time series. In panels (b) and (d), the brown dotted line in the FFT spectra denotes the 11-year cycle period for reference. The spectral power corresponding to the dominant period decreases during the grand minimum phase compared to the regular activity phase. We also observe a shift in the dominant period (shifts towards the lower period) during the grand minimum in our simulations. This behaviour is also present in the reconstructed OSF time series.}
    \label{fig:4a}
\end{figure*}

\begin{figure*}
	\includegraphics[width=0.9\textwidth]{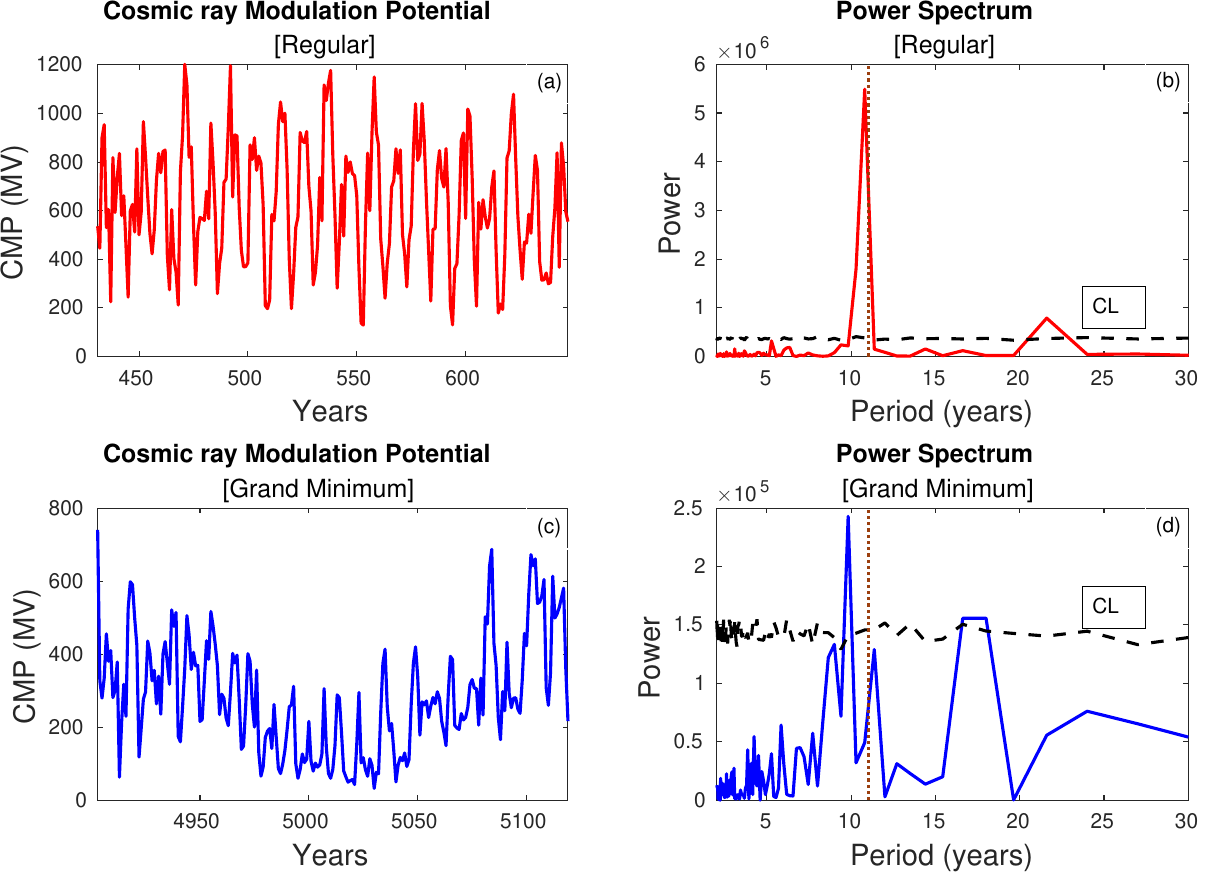}
    \caption{Spectral analysis on the simulated cosmic ray modulation potential time series. (a) The red curve shows the time series of cosmic ray modulation potential for a regular activity phase (where there are finite sunspot eruption proxies on the solar surface) from year 432 to 648. (b) The red curve shows the power spectrum of the time series. We have chosen the FFT window size to be 1-year for our calculations. The overplotted black dashed curve denotes the upper 95th percentile of the FFT spectra of 1000 re-sampled cosmic ray modulation potential for the regular activity phase. This indicates the confidence level (CL). (c) The blue curve denotes the modeled modulation potential time series for one of the grand minimum episodes (year 4904 to 5120). (d) The power spectrum of the cosmic ray modulation potential for the grand minimum phase. The confidence level is plotted in the black dashed line for this phase as well. Eleven-year periodicity is shown with a brown dotted curve on the FFT spectra. The power stored against the dominant period decreases during the grand minimum phase as compared to the regular phase.}
    \label{fig:4b}
\end{figure*}

The evolution of heliospheric parameters i.e. OSF and cosmic ray modulation potential also follow the solar activity cycle. In order to explore the periodicities involved in the time series of OSF and cosmic ray modulation potential we perform spectral analysis. Since there are multiple occurrences of grand minimum episodes in our long term simulation, we choose the longest grand minimum phase for our analysis which is from Year 4904 to 5120 (216 years). For comparison, we also select a regular activity phase of the same length (from Year 432 to 648). We use Fast Fourier Transform (FFT) with an annual cadence for deriving the power spectrum of OSF and cosmic ray modulation potential time series. Fig. \ref{fig:4a} and \ref{fig:4b} show the time series and the power spectrum during a regular activity phase and a grand minimum phase for OSF and cosmic ray modulation potential respectively.

In Fig. \ref{fig:4a}, for the power spectrum of simulated OSF, the dominant period during a regular phase is found to be 11 years (denoted by the brown dashed curve in Fig. \ref{fig:4a} (b)). However, for the grand minimum phase, we find a shift in the period towards a lower value with a significant reduction in power as compared to the regular phase (Fig. \ref{fig:4a} (d)). For calculating the confidence levels we randomize the sample 1000 times and perform FFT on each of the realizations. The upper 95th percentile for each period of these 1000 power spectra is defined as our confidence level (CL) which is denoted by the dashed black curve for all the spectral analyses. For the simulated cosmic ray modulation potential, the power spectrum shows a dominant period of 11 years during the regular phase (Fig. \ref{fig:4b} (b)). However, for a grand minimum phase, the dominant period again shifts towards a lower value with lower power than the regular phase (Fig. \ref{fig:4b} (d)). The dominant peaks are identified by the CL curve on the power spectrum.

\begin{figure*}
	\includegraphics[width=0.9\textwidth]{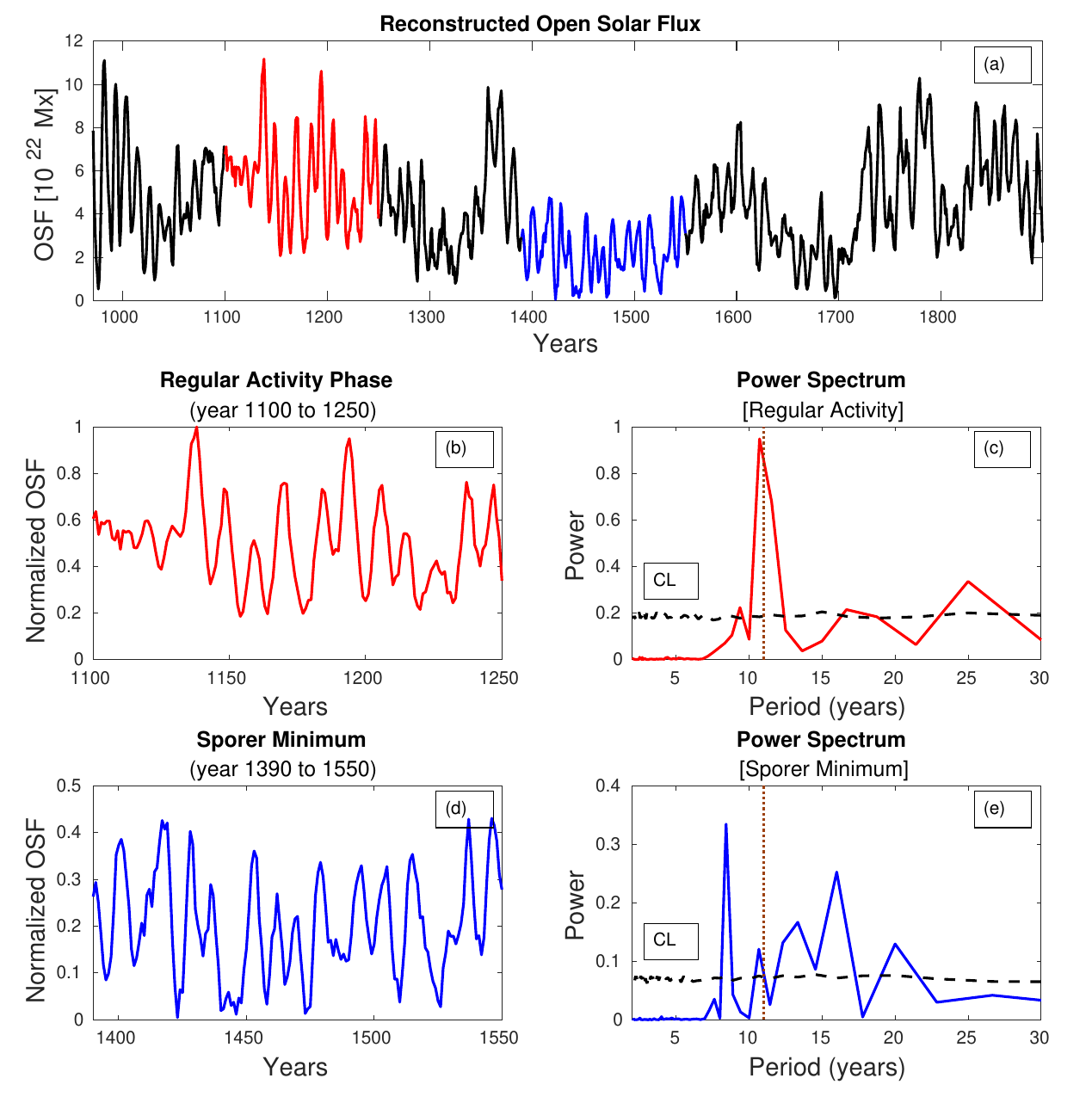}
    \caption{Fourier analysis of the reconstructed OSF. Reconstructed OSF data is obtained from \citet{Usoskin2021}. (a) The time series of the OSF from year 971 to 1899 is shown in the solid black curve with the red-colored region showing a regular activity phase (from year 1100 to 1250) and blue-colored region -- spanning from year 1390 to 1550 -- the Sp\"orer minimum (one of the observed grand minimum episodes). (b) Normalized OSF for the regular activity phase is shown in the solid red curve. (c) The power spectrum of the regular activity phase is denoted by the solid red curve. Here the brown dotted line denotes an 11-year period for reference. (d) Normalized OSF for Sp\"orer minimum phase is plotted in the solid blue curve. Fourier power spectrum corresponding to the Sp\"orer minimum phase is shown in panel (e). The solid black curve in panels (c) and (e) depicts the upper 95th percentile of the FFT spectra for 1000 re-sampled OSF time series. This determines the confidence level (CL). The power stored in the dominant cycle period decreases significantly during the grand minimum phase as compared to the regular activity phase. We find a similar trend in our modeled OSF as well.}
    \label{fig:5}
\end{figure*}

For comparison, we also compute the power spectrum using the observed 1000 year reconstructed OSF time series by \citet{Usoskin2021}. The OSF is sampled at an annual cadence in this data-set. For the grand minimum phase, we choose Sp\"orer minimum (denoted by blue color in Fig. \ref{fig:5} (a) and (d)) which starts from year 1390 and ends at 1550. In order to compare with a regular phase we choose a segment from year 1100 to 1250 which is a regular activity phase. Historical grand minimum time periods from the reconstructed data are reported in \citet{Usoskin2016AA}. The power spectrum of the reconstructed OSF during the regular phase shows a dominant period of 11 years (Fig. \ref{fig:5} (c)). On the other hand, the power in the dominant period during the grand minimum phase is lower than in the regular phase. Therefore the observations also show a similar shift in the power of the dominant period towards a lower value. This trend is independently confirmed by \citet{Saha2022} who focused only on exploring the persistence of weak magnetic activity during grand minima phases. This demonstrates the qualitative consistency of spectral power distribution shifts in the observed OSF reconstruction and our long-term simulations.

\section{Concluding Discussions}

To summarize, we have simulated a 6000-year long solar activity time series covering grand maxima, grand minima and regular activity phases utilizing a stochastically forced 2.5D kinematic solar dynamo model, and potential field extrapolation technique. We also calculate the open solar flux and the cosmic ray modulation potential to understand the impact of solar activity on the state of the heliosphere. The stochastically forced dynamo results in self-consistent generation of grand minima, maxima and regular activity phases; these variations are manifested on the surface which governs the evolution of coronal magnetic fields and in turn modulates the heliospheric parameters e.g., OSF and cosmic ray modulation potential. 

We find that the global configuration of the coronal magnetic fields changes during these different phases. The open flux that drives the solar wind into the interplanetary medium reduces during grand minima phases. We also find closed field regions distributed across all latitudes. When the height of the source surface in the PFSS model is lowered these closed magnetic field lines persist very low down close to the surface. Towards the end of the Maunder minimum (which is a grand minimum phase), a few solar eclipse paintings and simulations show a structure-less large scale solar atmosphere \citep{Riley2015,Lockwood2021A&G,Hayakawa2021}. We surmise that less solar open field lines (and lower open flux) during grand minima phases -- resulting from the surface field distribution governed by the solar dynamo output -- culminates in a weaker driving of the solar wind. The cumulative effect of this is an overall enhancement in the cosmic ray flux arriving at Earth during weak solar activity phases.


While explicit solar wind simulations are beyond the scope of this current paper, such fluctuations are reported in \citet{owens2017} where the authors reconstruct the solar wind for the past four centuries. In these reconstructions, the overall solar wind speed is indeed significantly reduced during grand minimum phases and is higher during the regular phases. In this context, we point out the study by \citet{Pinto2011ApJ} who explore the physics of direct coupling between the dynamo and the solar wind. Understanding the physics of this coupling is essential to causally connect dynamics from the solar interior to its atmosphere 
\citep{Perri2018,Perri2020JSWSC,Perri2021ApJ}.

The source surface is expected to vary with solar activity levels. While a self-consistent treatment of this is only possible in dynamical models of coupled corona and interior, we have performed a few heuristic numerical experiments to study the impact of varying the local source surface in an ad hoc manner which indeed has an impact on the structuring of the corona and the resultant open flux and cosmic ray modulation potential.

Cosmic ray flux is often used to reconstruct past solar activity. Reconstructions of past solar activity cycle based on cosmogenic isotope data encompass multiple occurrences of grand maxima, grand minima and regular activity phases \citep{Usoskin2007,Wu2018}. Variations present in the reconstructed data is suggestive of fluctuations in the solar forcing. During grand minima phases, there is an observed reduction in the dominant power and a shift to a lower cycle period. Our long-term dynamo simulations covering diverse phases of solar activity are in qualitative agreement with this; another study focusing on solar cycle dynamics during grand minima phases lends independent support to these results \citet{Saha2022}.

In conclusion, based on a 6000 year kinematic dynamo simulation and potential field extrapolations of the coronal field forced by the dynamo generated magnetic fields, we find significant changes in the coronal structure and heliospheric forcing parameters such as the open flux and cosmic ray modulation potential during distinct phases of solar activity. Our findings are qualitatively consistent with long term reconstructions of solar open flux. Our study provides a theoretical basis for establishing causality between the solar dynamo mechanism and the long-term forcing of the state of the heliosphere. It also lays a firm foundation for reconstruction of long-term (millennial scale) solar activity based on cosmogenic isotopes. 

\section*{Acknowledgements}

DN acknowledges fruitful exchanges at the Workshop on ``Solar and Stellar Dynamos: A New Era'' sponsored by the International Space Science Institute, Bern, where the idea of this work was initiated. The authors are thankful to Chitradeep Saha and Shaonwita Pal for helpful discussions. We acknowledge an anonymous referee for useful comments. SD acknowledges PhD fellowship from the DST-INSPIRE program of the Government of India. IU acknowledges partial support by the Academy of Finland (Project ESPERA No.~321882) and a visiting fellowship at ISSI (Bern). The Center of Excellence in Space Sciences India (CESSI) is funded by IISER Kolkata, Ministry of Education, Government of India.

\section*{Data Availability}
We have used 1000-year open solar flux data available at VizieR online data catalogue \citep{Usoskin2021}. We have also used the decadal averaged reconstructed sunspot number data available at VizieR online data catalogue \citep{Wu2018}. Data from our simulations will be shared upon reasonable request to the corresponding author.



\bibliographystyle{mnras}
\bibliography{reference}



\bsp	
\label{lastpage}
\end{document}